\documentclass[useAMS,usenatbib,onecolumn,referee]{mn2e}

\usepackage{graphicx}
\usepackage{natbib}
\usepackage{bm}

\title[The magnetohydrodynamics model of twin kilohertz QPOs]{The magnetohydrodynamics model of twin kilohertz QPOs in LMXBs}
\author[Shi \& Li]{ Changsheng Shi$^{1,2}$\thanks{E-mail:
scs1217@gmail.com} and Xiang-Dong Li$^{1}$\\
$^{1}$Department of Astronomy, Nanjing University, Nanjing 210093, China;\\
$^{2}$College of Material Science and Chemical Engineering, Hainan
University, Hainan 570228, China}

\begin{document}

\date{Accepted ??; Received ??; in original form
 ??}

\pagerange{\pageref{firstpage}--\pageref{lastpage}} \pubyear{2008}

\maketitle

\label{firstpage}

\begin{abstract}
We suggest an explanation for the twin kilohertz quasi-periodic
oscillations (kHz QPOs) in low-mass X-ray binaries (LMXBs) based on
magnetohydrodynamics (MHD) oscillation modes in neutron star
magnetospheres. Including the effect of the neutron star spin, we
derive several MHD wave modes by solving the dispersion equations,
and propose that the coupling of the two resonant MHD modes may lead
to the twin kHz QPOs. This model naturally relates the upper, lower
kHz QPO frequencies with the spin frequencies of the neutron stars,
and can well account for the measured data of six LMXBs.
\end{abstract}


\section{Introduction}
\label{intro}

The fastest variability components in X-ray binaries, the kilohertz
quasi-periodic oscillations (kHz QPOs), have been detected in about
thirty neutron star low-mass X-ray binaries (NS LMXBs) since first
discovered with Rossi X-ray Timing Explorer in 1996 (see van der
Klis 2006 for a review). Twin kHz QPOs appeared simultaneously in
about twenty NS LMXBs. These QPOs are thought to reflect the motion
of matter at the inner edge of an accretion disc around the NS.
Early observations showed that the frequency separation ($\Delta
\nu$) of the twin kHz QPOs is close to the NS spin frequency
($\nu_s$) (e.g. Strohmayer et al. 1996; Ford et al. 1997),
suggesting a beat-frequency explanation (Miller, Lamb \& Psaltis
1998). However, the more detailed measurements revealed that $\Delta
\nu$ is generally inconsistent with a constant value of $\nu_s$ but
varying with the upper ($\nu_2$) or lower ($\nu_1$) frequency of the
twin kHz QPOs (van der Klis et al. 1997; M\'endez et al. 1998,
1999). Stella \& Vietri (1999) propose a relativistic precession
model which predicts a changing peak separation of the twin kHz
QPOs. The upper and the lower kHz QPO frequencies are identified
with the Keplerian frequency of the rotational plasma flow at the
inner edge of the disc and the periastron precession frequency of
the orbit, respectively. In this model, the NS spin frequency plays
no role in setting up the frequencies of the kHz QPOs, but a massive
($\sim 2M_\odot$) NS is usually required to match the observations.
Abramowicz et al. (2001, 2003) suggest that the twin kHz QPOs can be
explained by a nonlinear resonance in the epicyclic motion in the
accretion disc, leading to the $3 : 2$ ratio  of the upper and lower
frequencies, although whether there is an intrinsically preferred
ratio between the $\nu_2$ and $\nu_1$ is controversial (Belloni et
al. 2007).

The subtle feature of twin kHz QPOs is that $\Delta \nu$ is variable
but seems to be around either $\nu_s$ or $\nu_s/2$ (Smith, Morgan \&
Bradt 1997; Wijnands \& van der Klis 1997; Markwardt, Strohmayer \&
Swank 1999; Chakrabarty et al 2003; Wijnands et al. 2003; Barret,
Boutelier \& Miller 2008; see, however, Yin et al. 2007; M\'endez \&
Belloni 2007). This has motivated some scenarios of kHz QPOs taking
into account the effect of the NS spin. Osherovich \& Titarchuk
(1999) suggest that kHz QPOs could be modeled as oscillations of the
blobs thrown into magnetosphere from the inner edge of the accretion
disc. The lower kHz QPO frequency $\nu_1$ is identified as the
Keplerian frequency in the disc and the upper frequency $\nu_2$ the
hybrid frequency about the Keplerian frequency and the NS spin
frequency. Lee, Abramowicz \& Klu\'niak (2004) show that the kHz
QPOs may be attributed to forcing of epicyclic motions in the
accretion disc by the NS, which induces resonance at selected
frequencies when the frequency separation $\Delta \nu$ is equal to
$\nu_s$ or $\nu_s/2$. Li \& Zhang (2005, see also Zhang 2004)
present an alternative interpretation for the origin of the twin kHz
QPOs by considering the interaction between the NS magnetic field
and the surrounding accretion disc, which gives rise to MHD loop
oscillations at the inner edge of the disc with the frequencies
depending on the spin frequencies.

In this paper, we propose a  model to explain kHz QPOs in NS LMXBs
based on the interaction of accreting plasma with the NS
magnetosphere. This model is partially based on the interpretation
of Zhang (2004) and Rezania \& Samson (2005). In the latter work it
was argued that distortion of the NS magnetosphere by the infalling
plasma of the Keplerian accretion flow can excite resonant shear
Alfv\'en waves in a region of enhanced density gradients, where
accretion material flows along the magnetic field lines in the
magnetosphere. The major difference between our model and Rezania \&
Samson's (2005) is that we have included the effect of the gravity
and spin of the NS on the field line resonances.

This paper is organized as follows. In section 2 we introduce the
basic physical model and derive the relation between $\nu_s$,
$\nu_1$ and $\nu_2$. In section 3 we compare the theoretical
relations with the observational data of six sources (4U 0614$+$09,
4U 1608$-$52, 4U 1636$-$53, 4U 1728$-$34, 4U 1915$-$05, and XTE
1807$-$294). In section 4 we summarized our results and discuss
their possible implications.

\section{ THE MODEL}

\label{code}

In LMXBs the plasma from the donor star accretes onto the NS via an
accretion disc. Material in the disc firstly rotates in a Keplerian
motion, then corotates with the magnetosphere after it is trapped
completely by the NS magnetic field at the magnetospheric radius,
and finally flows along the field lines to the polar cap. Some
resonant modes may be excited by the perturbations at the
magnetospheric radius when the plasma begins to corotate with the
magnetosphere (Osherovich \& Titarchuk 1999; Lee et al. 2004; Zhang
2004; Rezania \& Samson 2005). We consider the QPOs as a modulation
effect of the MHD waves which are produced at the magnetospheric
radius, and the coupling of the two resonant MHD modes may lead to
the twin kHz QPOs in the power spectrum.

We consider the MHD equations in a frame of reference corotating
with the NS (shown in Fig.~1), written as follows (Landau \&
Lifshitz 1976)
\begin{equation}
\label{eq1} \rho \frac{\mbox{d}{{\bm v}}}{\mbox{d}t} = -\nabla P + {
{\bm J}}\times { {\bm B}} + 2\rho {\bm v}\times {\bm\Omega } + \rho
{\bm \Omega }\times ({\bm r}\times {\bm \Omega }) - \rho
\textstyle{{GM} \over {r^3}}{\bm r},
\end{equation}
\begin{equation}
\label{eq2} \frac{\partial{\bm{B}}}{\partial t} = \nabla \times
({\bm v}\times {\bm B}) = ({\bm B} \cdot \nabla ){\bm v} - ({\bm v}
\cdot \nabla ){\bm B} - (\nabla \cdot {\bm v}){\bm B},
\end{equation}
\begin{equation}
\label{eq3} {\partial{\rho} \over \partial t} +\nabla\cdot({\rho
{\bm v}})=0,
\end{equation}
\begin{equation}
\label{eq4} P\rho^{-\gamma}={\rm const},
\end{equation}
where ${\bm v}$ is the plasma velocity, ${\bm J}$ electric current,
${\bm B}$ magnetic field, $\bm r$ the displacement from the center
of the NS to the plasma, $\rho$ plasma density, $P$ barometric
pressure, $\gamma$ adiabatic index, $G$ gravitational constant, $M$
and ${\bm \Omega}$ the mass and the angular velocity of the
NS\footnote{Actually ${\bm \Omega}$ is the angular velocity of the
NS magnetosphere, which could be slightly deviate from that of the
NS.}, respectively. The third, fourth and fifth terms on the rhs of
Eq.~(1) represent the Coriolis force, the centrifugal force and the
gravity, respectively.

Observationally the accretion rates in LMXBs change on a timescale
of $\sim10^3-10^4 s$. This is much more than the relaxation (or
dynamic) timescale of the plasma at the inner disc radius
($\sim10^{-2}-10^{-3} s$). So we can approximate the plasma to be in
an equilibrium state, which is subject to small perturbations, as
discussed in Benz (2002).By use of the current expression,
\begin{equation}
\label{eq5} {{\bm J}} = \textstyle{1 \over \mu }\nabla \times {{\bm
B}},
\end{equation}
where $\mu$ is is vacuum magnetic conductivity, Eqs.~(1)-(3) can be
transformed to be
\begin{equation}
\label{eq6} \rho _0 \textstyle{{\partial { {\bm v}}_{ {\bm 0}} }
\over {\partial t}} + ({{\bm v}_{\bm 0}}\cdot \nabla){{\bm v}_{\bm
0}} = -\nabla P_0 + \textstyle{1 \over \mu }(\nabla \times {{\bm
B}}_{ {\bm 0}} )\times { {\bm B}}_{ {\bm 0}} +2\rho _0 {{\bm v}}_{
{\bm 0}} \times { {\bm \Omega }} +\rho _0 \Omega ^2{ {\bm r}}_{ {\bm
0}} - \rho _0 ({ {\bm \Omega }} \cdot { {\bm r}}_{{\bm 0}} ){{\bm
\Omega }} - \rho _0 \textstyle{{GM} \over {r_0 ^3}}{ {\bm r}}_{{\bm
0}},
\end{equation}
\begin{equation}
\label{eq7} \frac{\partial { {\bm B}}_{ {\bm 0}} }{\partial t} =
({{\bm B}}_{ {\bm 0}} \cdot \nabla ){ {\bm v}}_{ {\bm 0}} - ({ {\bm
v}}_{ {\bm 0}} \cdot \nabla ){{\bm B}}_{ {\bm 0}} - (\nabla \cdot
{{\bm v}}_{ {\bm 0}} ){{\bm B}}_{ {\bm 0}},
\end{equation}
\begin{equation}
\label{eq8} {\partial{\rho_0} \over \partial t}
+\nabla\cdot({\rho}{{\bm v_0}})=0,
\end{equation}
where the subscript 0 denotes variables in the equilibrium state.
The initial relative velocity ($ {\bm {v_0} }$) is equal to zero in
the corotating reference system and can be expressed as ${{ {\bm
\Omega }}\times { {\bm r}}_0 }$ in the inertial reference system.

Now we consider the MHD equations for the plasma subject to small
perturbations,
\begin{equation}
\label{eq9} \rho _0 \textstyle{{\partial \hat{\bm v}} \over
{\partial t}} + ({\hat{\bm v}}\cdot \nabla){\hat{\bm v}} = -\nabla
\hat{P} + \textstyle{1 \over \mu }(\nabla \times \hat{\bm B})\times
\hat{\bm B} + 2\rho _0 \hat{\bm v}\times {{\bm \Omega }} + \rho _0
\Omega ^2\hat{\bm r} - \rho _0 ({\rm {\bm \Omega }} \cdot \hat{\bm
r}){ {\bm \Omega }} - \rho _0 \textstyle{{GM} \over {r_0
^3}}\hat{\bm r},
\end{equation}
\begin{equation}
\label{eq10} \frac{\partial \hat{\bm B}}{\partial t} =  ({ \hat{\bm
B}} \cdot \nabla )\hat{\bm v} - ({ \hat{\bm v}} \cdot \nabla ){
\hat{\bm B}} - (\nabla \cdot \hat{\bm v})\hat{\bm B}.
\end{equation}
\begin{equation}
\label{eq11} {\partial{\hat{\rho}} \over \partial t}
+\nabla\cdot(\hat{{\rho}}\hat{{{\bm v}}})=0,
\end{equation}
\begin{equation}
\label{eq12} \hat{P}\hat{\rho}^{-\gamma}= P_0\rho_0^{-\gamma},
\end{equation}
where $\hat{\bm v} = {{\bm v}}{ }_{ {\bm 0}} + { {\bm v}}_{ {\bm s}}
= { {\bm v}}_{{\bm s}}$, $\hat{\bm B} = { {\bm B}}{ }_{ {\bm 0}} +
{{\bm B}}{ }_{{\bm s}}$, $\hat{\bm r} = {{\bm r}}{ }_{{\bm 0}} +
{{\bm r}}{ }_{{\bm s}}$, $\hat{\rho} = \rho_0 + \rho_s$, $\hat{P} =
P_0 + P_s$ with the subscript s denoting the perturbed quantities
($v_s \ll \left| {{ {\bm \Omega }}\times { {\bm r}}_0 }\right|$,
$B_s\ll B{ }_0$, $r_s \ll r{ }_0$, $\rho_s\ll \rho_0$, $P_s\ll P_0$)
and with the superscript $\hat{}$ the variables after the
disturbance. Combining Eqs.~(6)-(12) we get the equations about the
perturbed quantities in the first order approximation,
\begin{equation}
\label{eq13} \begin{array}{lll}
 \rho _0 \frac{\partial { {\bm v}}_{ {\bm s}} }{\partial t}&
 =  & -{\gamma P_0\over {\rho_0}}{\bigtriangledown\rho_s} + \textstyle{1 \over \mu }[(\nabla \times {{\bm B}}_{{\bm 0}}
)\times { {\bm B}}_{{\bm s}} + (\nabla \times { {\bm B}}_{ {\bm s}}
)\times { {\bm B}}_{ {\bm 0}} ] + 2\rho _0 { {\bm v}}_{ {\bm s}}
\times { {\bm \Omega }} +
 \rho _0 \Omega ^2{ {\bm r}}_{{\bm s}}\\
  & & - \rho _0 ({ {\bm \Omega }}
\cdot { {\bm r}}_{ {\bm s}} ){ {\bm \Omega }} - \rho _0
\textstyle{{GM} \over {r_0 ^3}}{ {\bm r}}_{ {\bm s}},
 \end{array}
\end{equation}
\begin{equation}
\label{eq14} \frac{\partial { {\bm B}}_{ {\bm s}} }{\partial t} = ({
{\bm B}}_{{\bm 0}} \cdot \nabla ){{\bm v}}_{{\bm s}} - (\nabla \cdot
{{\bm v}}_{ {\bm s}} ){ {\bm B}}_{ {\bm 0}},
\end{equation}
and
\begin{equation}
\label{eq15} {\partial{\rho_s} \over \partial t}
+{\rho_0}\nabla\cdot{{\bm v_s}}=0.
\end{equation}
 Differentiating Eq.~(13) and substituting $\nabla (\bm {B_{0}}  \cdot \bm {B_{s}}  ) = (\bm {B_{0}} \cdot
\nabla )\bm {B_{s}}  + (\bm {B_{s}} \cdot \nabla )\bm {B_{0}} + \bm
{B_{s}} \times (\nabla \times \bm {B_{0}}  ) + \bm {B_{0}} \times
(\nabla \times \bm {B_{s}}  )$ into it give
\begin{equation}
\label{eq16}
\begin{array}{lll}
 \rho _0 \frac{\partial ^2{ {\bm v}}_{{\bm s}} }{\partial t^2}
& = & - {{\gamma P_0\over {\rho_0}}\frac{\partial }{\partial
t}({\bigtriangledown\rho_s}}) + \textstyle{1 \over \mu
}\frac{\partial }{\partial t}[({ {\bm B}}_{ {\bm 0}} \cdot \nabla ){
{\bm B}}_{ {\bm s}} + ({ {\bm B}}_{ {\bm s}} \cdot \nabla ){ {\bm
B}}_{ {\bm 0}} - \nabla ({{\bm
B}}_{ {\bm 0}} \cdot { {\bm B}}_{{\bm s}} )]\\
& & + 2\rho _0 \textstyle{\partial \over {\partial t}}({ {\bm v}}_{
{\bm s}} \times {{\bm \Omega }}) + \rho _0 \Omega ^2{{\bm v}}_{ {\bm
s}} - \rho _0 ({ {\bm \Omega }} \cdot { {\bm v}}_{{\bm s}} ){ {\bm
\Omega }} - \rho _0 \textstyle{{GM} \over {r_0 ^3}}{{\bm v}}_{ {\bm
s}}.
\end{array}
\end{equation}

We assume that (1) the accretion disc is infinitesimally thin, (2)
the magnetic moment and the spin of the NS are parallel to the $z$
axis, and normal to the disc, i.e., ${ {\bm B}}_{ {\bm 0}} =
(0,0,B_0 )$ and ${ {\bm \Omega}}=(0,0,\Omega )$ close to the inner
edge of the disc, and (3) the $x$ and $y$ axes are along the disc
plane, and the MHD wave is assumed to propagate in the $xoz$ plane,
i.e., the wave vector ${ {\bm k}}= (k\sin \theta ,0,k\cos \theta )$,
where $\theta $ is the angle between the $z$ axis and $ {\bm k}$.
After carrying out Fourier transformation for Eqs.~(14)-(16) we get
the following dispersion equations,
\begin{equation}
\label{eq17} (1 + \frac{\Omega ^2}{\omega ^2} - \frac{\omega _k
^2}{\omega ^2} - \frac{k^2 V_A ^2}{\omega ^2} - \frac{k^2 c_s ^2
sin^2\theta}{\omega ^2})v_{sx} = \frac{k^2 c_s ^2 sin\theta
cos\theta}{\omega^2} v_{sz} + {\frac{2\Omega i}{\omega}}v_{sy},
\end{equation}
\begin{equation}
\label{eq18} (1 + \frac{\Omega ^2}{\omega ^2} - \frac{\omega _k
^2}{\omega ^2} - \frac{k^2 V_A ^2 cos^2\theta}{\omega ^2})v_{sy} =
-{\frac{2\Omega i}{\omega}}v_{sx},
\end{equation}
\begin{equation}
\label{eq19} (1 - \frac{\omega _k ^2}{\omega ^2} - \frac{k^2 c_s ^2
cos^2\theta}{\omega ^2})v_{sz} = \frac{k^2 c_s ^2 sin\theta
cos\theta}{\omega ^2}v_{sx},
\end{equation}
where $i$ is imaginary unit, $V_A$ ($= \sqrt {B_0 ^2 / \mu \rho _0
}$), $c_s$ ($= \sqrt {\gamma P_0 / \rho_0}$), $\omega _k$ ($= \sqrt
{GM / r_0 ^3}$), and $\omega$ are Alfv\'en velocity, acoustic
velocity, Keplerian angular velocity, and angular velocity of the
perturbation at $r_0$ respectively, and $v_{sx}$, $v_{sy}$, $v_{sz}$
are the three components of the perturbed quantity of the speed.
Equations (17)-(19) show that there exist three resonance MHD modes.
At the magnetospheric radius $r_0$, the magnetic energy density is
equal to the total kinetic energy density, i.e. $ B^2 / 8\pi = \rho
V_A^2/2 \simeq \rho V_k^2/2$ (Davidson \& Ostriker 1973; Ghosh et
al. 1977). Since the characteristic wavelength is in the same order
with the magnetospheric radius (Rezania {\&} Samson 2005), we then
have $kV_A \sim kV_K \sim V_K/r_0 = \omega_k $, or $kV_A = \eta
\omega _k$. Because the thermal pressure of the plasma might be
comparable with the magnetic pressure ($c_s  \sim V_A$) just inside
the magnetosphere (Miller et al. 1998), we also suppose $kc_s =
\lambda \omega _k$. Here both $\eta$ and $\lambda$ are  taken to be
constant for certain sources. Substitute these relations into
Eqs.~(17)-(19) we can get the resonant modes. Specifically when
$\theta = 0$, from Eqs.~(17) and (18) we can get $v_{sx} = \pm
iv_{sy}$, i.e. $v_{sx} e^{i{\bm k} \cdot {\bm r} - i\omega t} =
v_{sy} e^{i{\bm k} \cdot {\bm r} - i\omega t\pm i{\textstyle{{\pi}
\over {2}}}}$. Substituting this relation into the Eqs.~(17)-(19)
can give
\begin{equation}
\label{eq20} \omega_1 = \sqrt{1+ \eta ^2}\,\omega _k- \Omega,
\end{equation}
\begin{equation}
\label{eq21} \omega_2 = \sqrt{1+\lambda^2}\,\omega _k,
\end{equation}

\begin{equation}
\label{eq22} \label{eq20} \omega_3 = \sqrt{1+ \eta ^2}\,\omega _k+
\Omega,
\end{equation}
where the negative solutions are excluded. Similarly when $\theta
=\pi/2$,
\begin{equation}
\label{eq23} \omega_1^2 = \omega _k^2,
\end{equation}
\begin{equation}
\label{eq24} \omega_2^2 = \omega _k ^2 + \Omega ^2 + \frac{\omega
_k ^2}{2}(\eta ^2 + \lambda ^2) + \frac{\omega _k }{2}\sqrt {\omega
_k ^2(\eta ^2 + \lambda ^2)^2 + 8\Omega ^2(\eta ^2 + \lambda ^2 +
2)},
\end{equation}
\begin{equation}
\label{eq25} \omega_3^2 = \omega _k ^2 + \Omega ^2 + \frac{\omega
_k ^2}{2}(\eta ^2 + \lambda ^2) - \frac{\omega _k }{2}\sqrt {\omega
_k ^2(\eta ^2 + \lambda ^2)^2 + 8\Omega ^2(\eta ^2 + \lambda ^2 +
2)}.
\end{equation}
Note that in the latter case the MHD wave  couldn't propagate  very
far in the accretion disc, so we consider the coupling modes that
propagate along the $z$ axis ($\theta=0$) as a more promising
explanation of the QPOs.

We first rule out the possibility of the $\omega_2$ mode as the
lower kHz QPOs, since Eq.~(21) requires that it should always higher
than $\Omega$ for stable accretion, which is contradicted with
observations. The coupling between $\omega_1$ and $\omega_3$ can
also be excluded, which implies a constant frequency separation. So
we suggest the upper and the lower kHz QPO frequencies be $\nu _2 =
\omega_2 / 2\pi$ and $\nu_1 = \omega_1 / 2\pi$. From Eqs.~(20) and
(21) we can get the following relation between the frequencies of
the upper and the lower kHz QPOs,
\begin{equation}
\label{eq26}  {\nu _2} = \sqrt{\frac{1 + \lambda^2}{1 +
\eta^2}}({\nu _1}+{\nu_s})
\end{equation}
where $\nu_s= \Omega / 2\pi$, or
\begin{equation}
\label{eq27}  \frac{\nu _2}{\nu_s} = \frac{1}{\sqrt {1 +
\varepsilon^2}}(\frac{\nu _1}{\nu_s}+1)\ ({\rm when}\ \eta >
\lambda),
\end{equation}
\begin{equation}
\label{eq28} \frac{\nu _2}{\nu_s} = \sqrt {1 + \delta^2}(\frac{\nu
_1}{\nu_s}+1)\ ({\rm when}\ \eta < \lambda),
\end{equation}
where $\varepsilon^2 = (\eta ^2 - \lambda ^2)/(1 + \lambda ^2)$ and
$\delta^2 = (\lambda ^2 - \eta ^2)/(1 + \eta ^2)$. Equations (27)
and (28) indicate that the twin kHz QPOs may be divided into two
groups, with the slope of the $\nu_2/\nu_s$ vs. $\nu_1/\nu_s$
relation either larger or smaller than 1. In the following we call
them the large slope coefficient sources (LSCS) and the small slope
coefficient sources (SSCS), respectively.

\section{COMPARISON WITH OBSERVATIONS}

We compare in Fig.~2 the $\nu_2/\nu_s$ vs. $\nu_1/\nu_s$ relations
obtained in last section with the observed kHz QPOs in six sources
4U 0614$+$09, 4U 1608$-$52, 4U 1636$-$53, 4U 1728$-$34, 4U
1915$-$05, XTE 1807$-$294, in which both the spin and twin kHz QPO
frequencies have been measured. The spin frequencies, disposed in
Table 1, are from van der Klis (2006), M\'endez {\&} Belloni (2007),
Yin et al. (2007), Altamirano et al. (2008), and their references.
For 4U 0614+09 we adopt the updated spin frequency of 415 Hz
(Strohmayer, Markwardt \& Kuulkers 2008). The dots with error bars
represent the measured values, and the solid lines stand for
theoretical relations. We distinguish the $\nu_2/\nu_s$ vs.
$\nu_1/\nu_s$ relations for SSCS and LSCS, and accordingly adopt
relation (27) to fit the data for 4U 0614$+$09, 4U 1608$-$52, 4U
1636$-$53, and 4U 1728$-$34, and relation (28) for the other two
sources, 4U 1915$-$05 and XTE 1807$-$294. For each source, the value
of $\varepsilon$ or $\delta$ for best fitting is also shown in the
figure. It is noted that a cluster of the values ($\sim 0.3-0.9$) of
$\varepsilon$ and $\delta$ can well reproduce the observed
relations. In the left panel of Fig.~3 we show the observed and
predicted relations for all of the six sources. In the right panel
we plot the relation between $\Delta\nu/\nu_s$ and $\nu_1/\nu_s$ by
use of the parameter $\varepsilon $ or $\delta$ that we have got.

For SSCS the peak separation of the twin kHz QPOs is less than the
spin frequency, i.e.,  $\Delta \nu - \nu _s = -( 1-1/\sqrt {1 +
\varepsilon ^2})(\nu _1 + \nu _s )< 0$, and decreases with $\nu_1$
or $\nu_2$; for LSCS the peak separation is more than  the spin
frequency, i.e., $\Delta \nu = (\sqrt {1 + \delta ^2} - 1)\nu _1 +
\nu _s > \nu _s$, and increases with the increasing $\nu_1$ or
$\nu_2$. In the former group, $\Delta\nu$ is around $\nu_s$ for 4U
0614$+$09 and 4U 1728$-$34, and $\nu_s/2$ for 4U 1608$-$52 and 4U
1636$-$53 (Miller et al. 1998; van der Klis 1997; Stella, Vietri \&
Morsink 1999; Lewin \& van der Klis 2006; M'endez \& Belloni 2007).

Our final note is that when the Alf\'ven speed is equal to the
acoustic speed of the plasma, i.e. $ \eta= \lambda$, Eq.~(26) will
recover to the expression in the sonic-point beat-frequency model
(Miller et al. 1998). In this case the peak separation is equal to
the spin frequency and almost invariant.

\section{DISCUSSION AND CONCLUSIONS}

In this paper we propose a resonant MHD model for the twin kilohertz
QPOs in LMXBs. The modes of the MHD waves vertical and parallel to
the accretion disc are derived, and the twin kHz QPOs frequencies
are identified with the frequencies of the two resonant modes. In
this model the twin kHz QPO frequencies are correlated with the spin
frequencies, and the separation frequencies also change with the QPO
frequencies. We show that the measured relations between $\nu_1$,
$\nu_2$, and $\nu_s$ can be accounted for with reasonable values of
the input parameters.

There are several spin-involved MHD models for kHz QPOs in the
literature. Osherovich \& Titarchuk (1999) suggest that kHz QPOs can
be explained as oscillations of large scale inhomogeneities (hot
blobs) thrown into the NS magnetosphere. Participating in the radial
oscillations with Keplerian frequency, such blobs are simultaneously
under the influence of the Coriolis force. The derived frequency
relation is $\nu_2^2=\nu_1^2+(2\nu_s)^2$, or
$(\nu_2/\nu_s)^2=(\nu_1/\nu_s)^2+4$, which is plotted in the dotted
curve in Fig.~4. The significant deviation from the measured data
indicates that this model is not successful for most of the LMXBs.
In their MHD loop oscillation model Li \& Zhang (2005) derive a
linear frequency relation, i.e., $\nu _2/\nu _s = \xi \nu _1/\nu _s
+ 1 $ where $\xi\sim 1$ is an input parameter. Here the upper kHz
QPO frequency is assumed to the Keplerian frequency and the lower
kHz QPO is identified as the principal fast kink mode of the
standing MHD waves along the toroidal field lines at the
magnetospheric radius. The relation is also plotted in Fig.~4 in
solid curves. A comparison with the measured data shows that the fit
is acceptable for all the six sources.

Rezania \& Samson (2005) propose a model for QPOs in LMXBs based on
oscillating magnetohydrodynamic modes in NS magnetospheres. They
argue that the interaction of the accretion disc with the
magnetosphere can excite resonant shear Alfv\'en waves in a region
of enhanced density gradients, where accretion material flows along
the magnetic field lines in the magnetosphere. The predicted
$\nu_2/\nu_1$ ratio is found to be independent of the NS spin
frequencies. The main difference between Rezania \& Samson (2005)
and this work lies in that Rezania \& Samson (2005) assume that the
strong gravity of the NS produces a converging flow which will hit
the star's magnetosphere in a large velocity, while we consider the
motion of the plasma is still mainly Keplerian, before they enter
the magnetosphere and corotate with the magnetosphere. We also
include the effect of the gravity and the rotation of the NS for the
trapped plasma, and find that the resulting wave frequencies relate
with both the spin frequency and the gravity.

So far we have assumed that the spin and magnetic axes of the NS are
aligned in LMXBs. However, the existence of persistent millisecond
pulsations in some LMXBs indicates that the magnetic axis is at
least somewhat tilted from the spin axis. In this case the magnetic
field is no longer homogeneous at a given radius in the accretion
disc, and the orbit of plasma in the inner region of the disc
becomes non circular. Furthermore, the stellar magnetic field can
induce disk warping and precession (e.g. Lai 1999, 2003), and
modulate the orbit and hence the QPO frequencies.  For X-ray
binaries the disc precession timescale is usually of tens of days to
years (cf. Wijers \& Pringle 1999 and references therein), which is
much longer than the duration of each QPO observation. Additionally
for accretion-powered millisecond pulsars the magnetic inclination
is likely to be very small (Lamb et al., 2008). For the above
reasons we expect that the change of the QPO frequencies induced by
oblique magnetic fields might be very small compared with the
uncertainties in the measured frequencies.

Our results indicate that the peak separation is always related to
the spin frequency. What's more, there seems to be a weak positive
correlation between the spin frequency and the parameter
$\varepsilon$ for SSCS, which can be described as
$\varepsilon=2.28(\pm 0.16)(\nu_s/1000\,{\rm Hz})- 0.53(\pm 0.08)$,
plotted in the left panel of Fig.~5. Substitute this relation into
$\Delta\nu = (1/\sqrt {1 + \varepsilon^2})(\nu_1 + \nu_s) - \nu_1 $
we get a trend of $\Delta\nu $ changing with the spin frequency, as
plotted in the right panel of Fig.~5. From the dark black curve to
the light gray curve $\nu_1$ changes from 1000 Hz to 100 Hz in a
step of 100 Hz. We find that when $\nu_s$ increases, $\Delta\nu$
varies from $\sim \nu_s/2$ to $\sim \nu_s$, and finally to $\sim
\nu_s/2$. The transitions occur at $\nu_s\sim 100$ Hz and 500 Hz,
respectively.

The accretion process can take place only when the magnetospheric
radius is less than the corotation radius (e.g. Ghosh {\&} Lamb
1979). In other words, the Keplerian frequency at the magnetosphere
radius should be more than the spin frequency of the NS if accretion
process can take place. Because of this there is a minimum value of
the lower frequency of the twin kHz QPOs in SSCS, $\nu _1 >
{(\sqrt{1 + \varepsilon^2} - 1)}{\nu _s }$. Besides, due to the fact
that the peak separation must be positive values we can get the
maximal value for the upper frequency, $\nu _2 < \nu _s/(\sqrt{1 +
\varepsilon ^2} - 1)$ for SSCS. These may serve as possible evidence
to testify this model with future measurements of kHz QPOs in LMXBs.

\section*{acknowledgements}

The authors thank the anonymous referee for the helpful suggestion
on the manuscript. This work was supported by the Natural Science
Foundation of China under grant numbers 10573010 and 10221001.

\clearpage

\begin{figure}
\begin{center}
 \includegraphics[width=2.36in,height=2.11in]{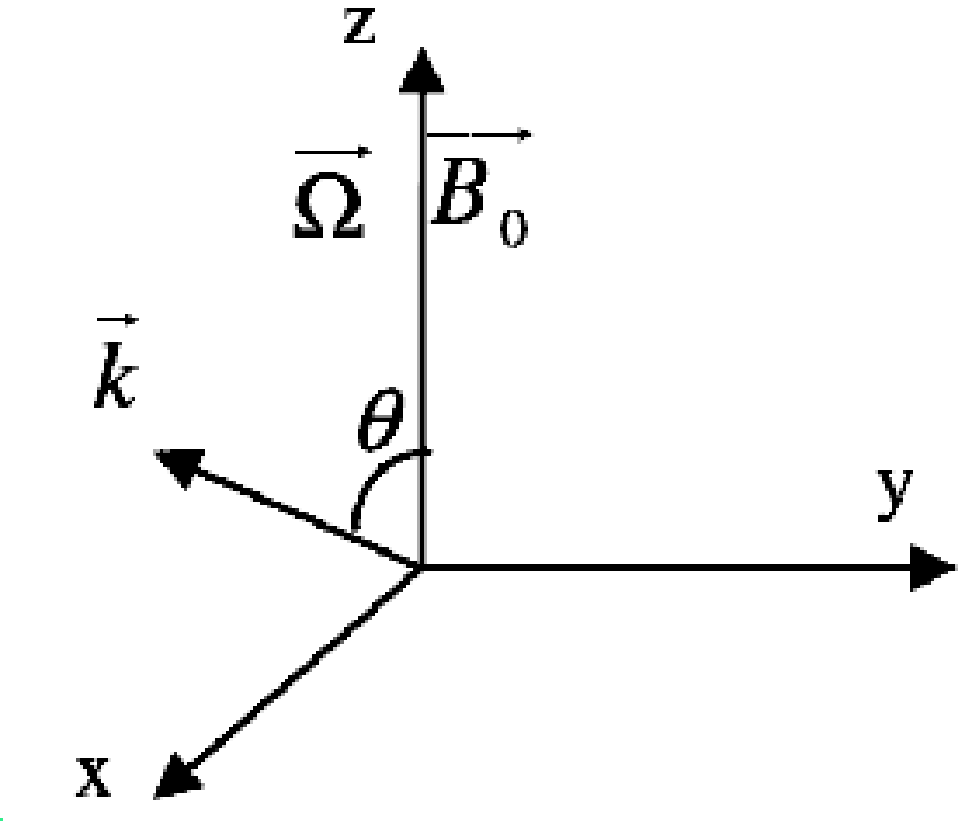}
 \end{center}
\caption{ The coordinate system centered at the magnetospheric
radius. The $x$ axis is along the radial direction, $y$ axis the
toroidal direction, and the $z$ axis is normal to the accretion
disc.} \label{fig1}
\end{figure}

\begin{center}
\begin{figure*}
\label{fig2}
 \includegraphics[width=2.9in,height=2.11in]{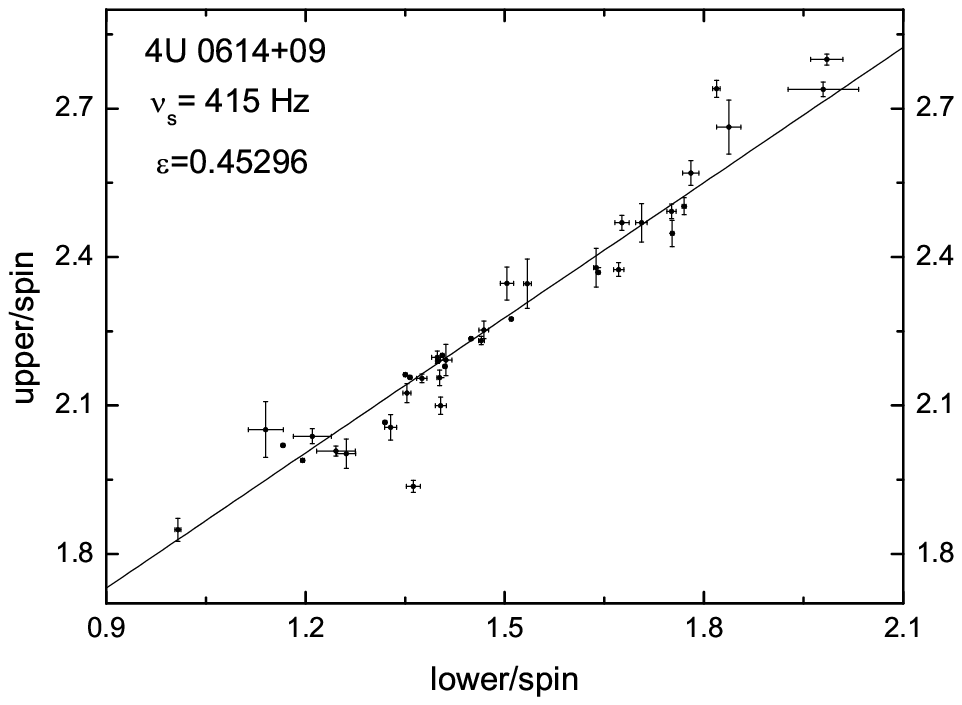}
 \includegraphics[width=2.9in,height=2.11in]{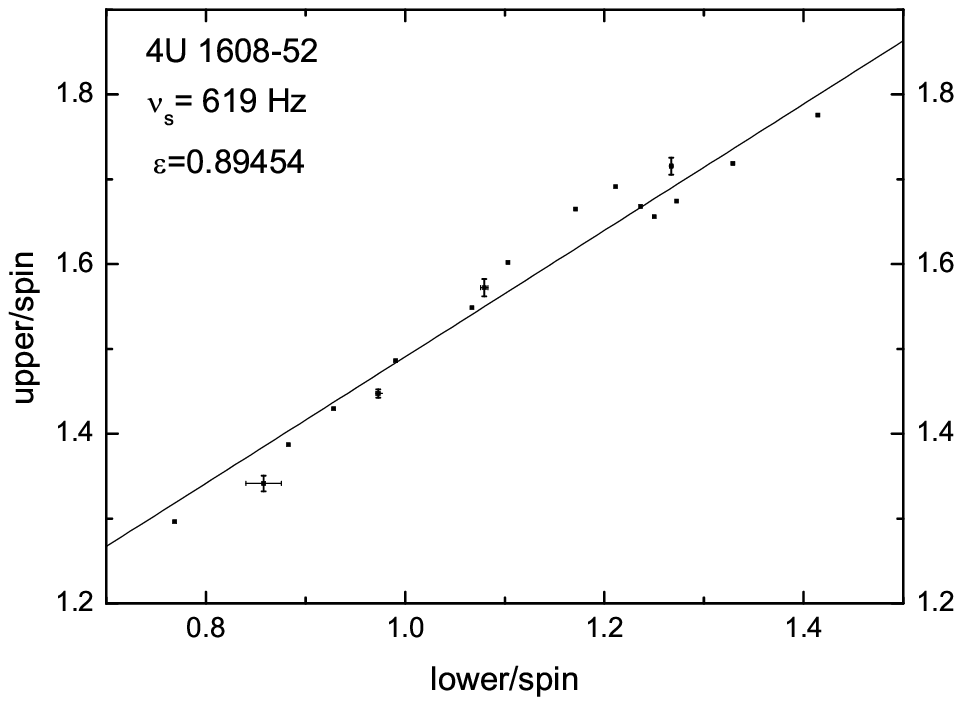}
 \includegraphics[width=2.9in,height=2.11in]{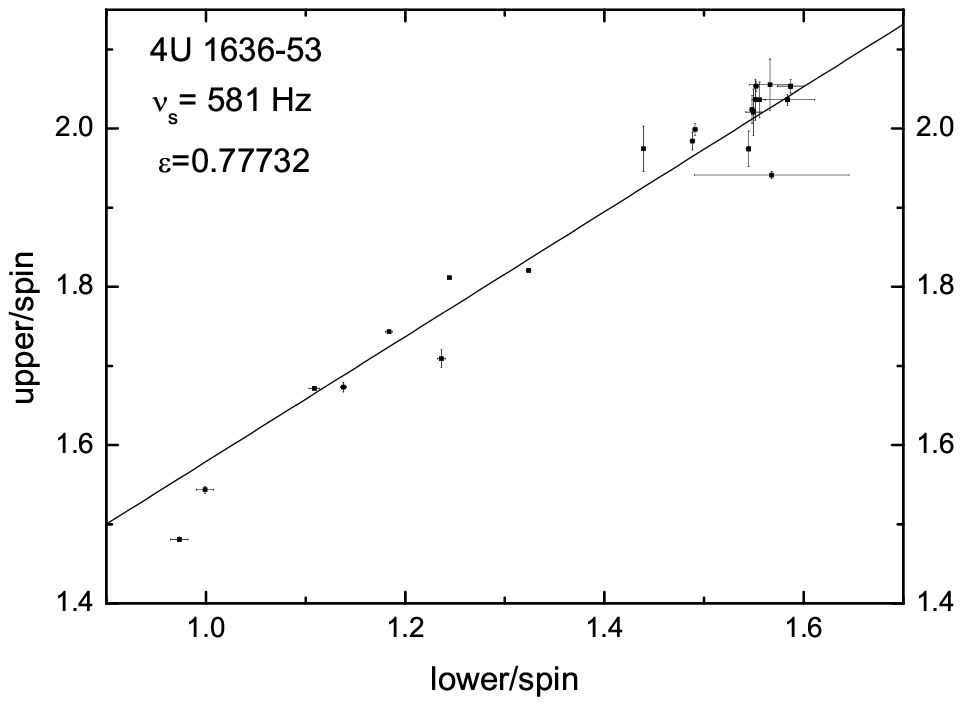}
 \includegraphics[width=2.9in,height=2.11in]{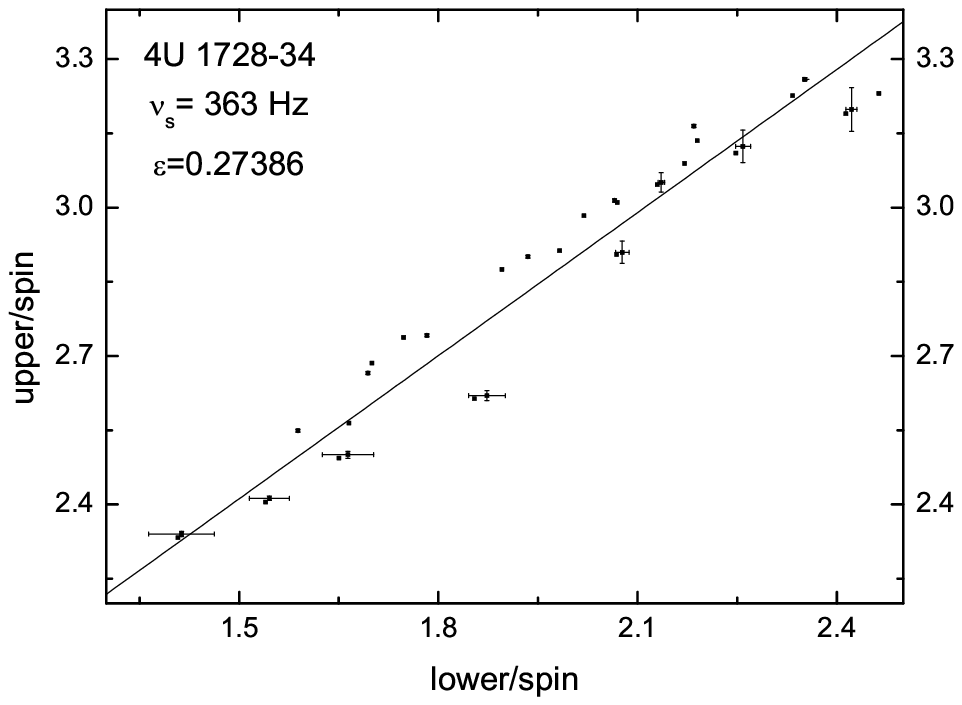}
 \includegraphics[width=2.9in,height=2.11in]{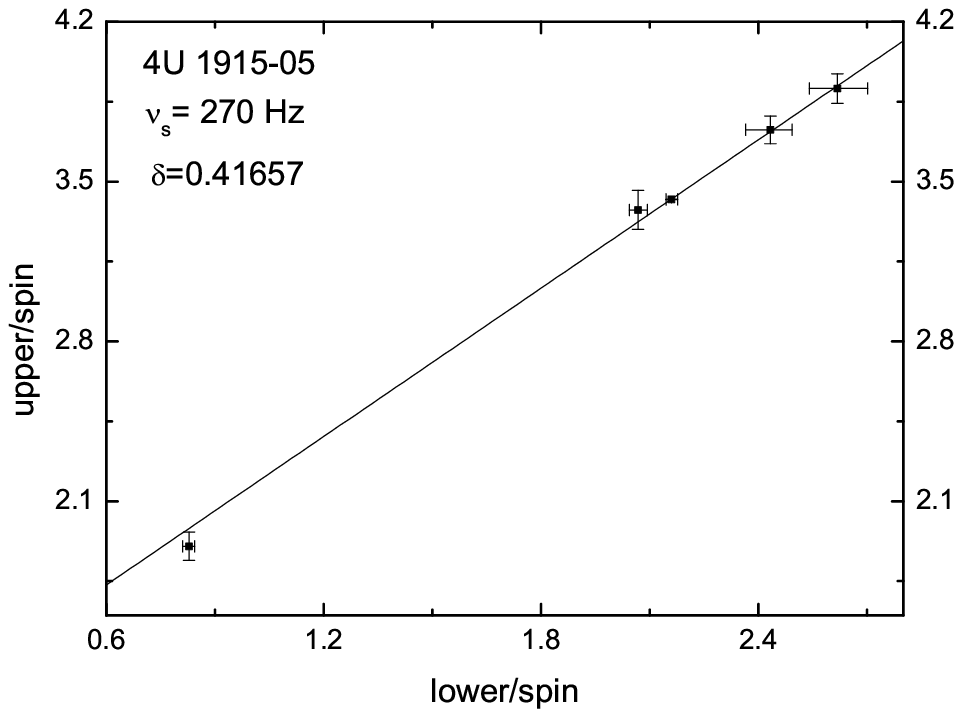}
 \includegraphics[width=2.9in,height=2.11in]{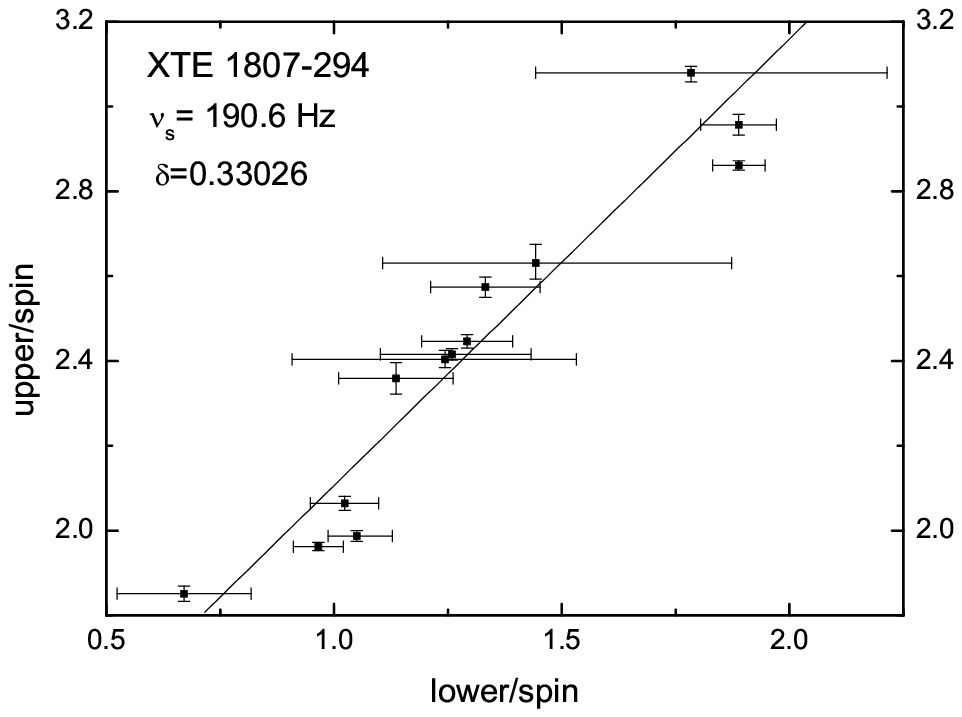}
\caption{ The relations between $\nu_2/\nu_s$ and $\nu_1/\nu_s$ for
the six sources (for the measured data of 4U 0614+09: van Straaten
et al. 2000; van Straaten et al. 2002; 4U 1608-52: van Straaten, van
der Klis \& M\'endez 2003; 4U 1636-53: Altamirano et al., 2008; Di
Salvo, M\'endez et \& van der Klis 2003; Jonker, M\'endez \& van der
Klis 2002; Wijands et al. 1997; 4U 1728-34: Migliari, van der Klis
\& Fender 2003; Di Salvo et al. 2001; Jonker, M\'endez \& van der
Klis 2000; 4U 1915-05: Boirin et al. 2000; XTE 1807-294: Linares et
al. 2005; Zhang et al. 2006).}
\end{figure*}
\end{center}

\begin{figure*}
\includegraphics[width=3.20in,height=2.35in]{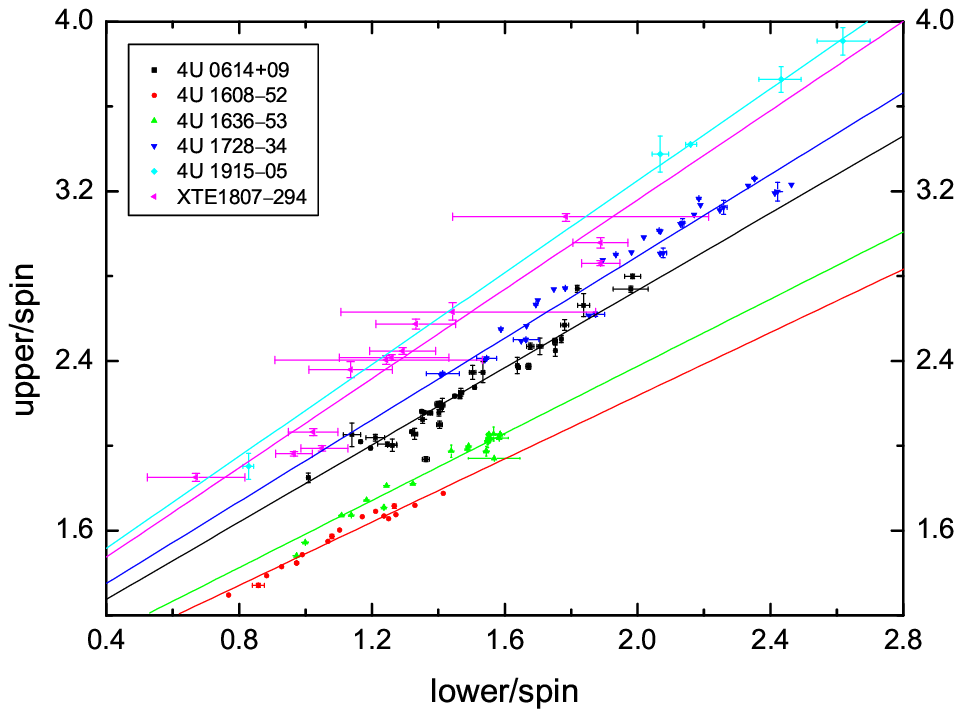}
\includegraphics[width=3.20in,height=2.35in]{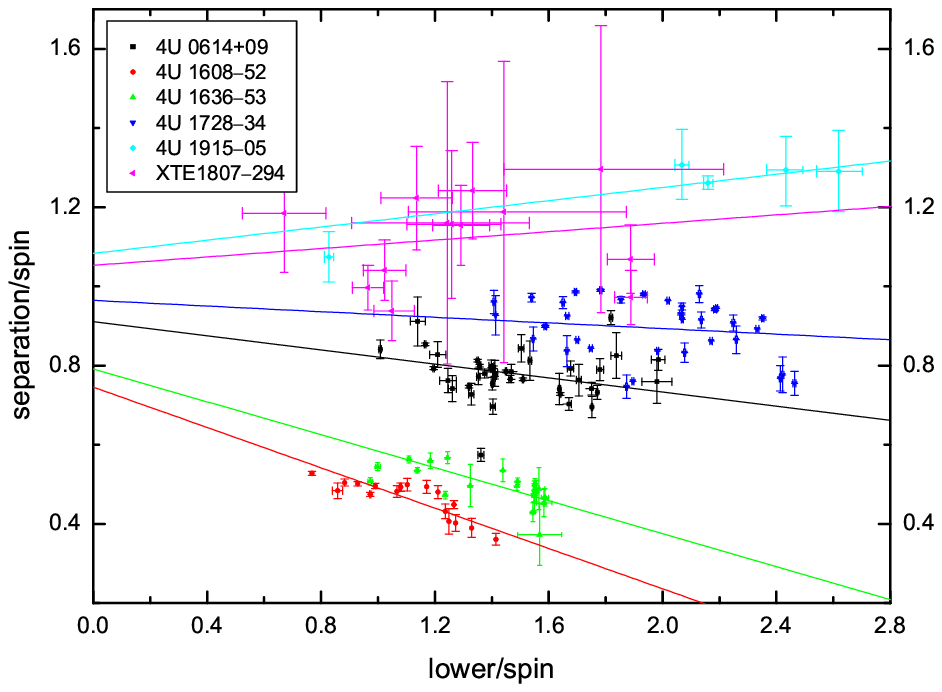}
\caption{ $Left$ The relation between $\nu_2/\nu_s$ and
$\nu_1/\nu_s$. $Right$ The relation between $\Delta\nu/\nu_s$ and
$\nu_1/\nu_s$.}\label{fig3}
\end{figure*}

\begin{center}
\begin{figure}
 \begin{center}
 \includegraphics[width=3.66in,height=2.85in]{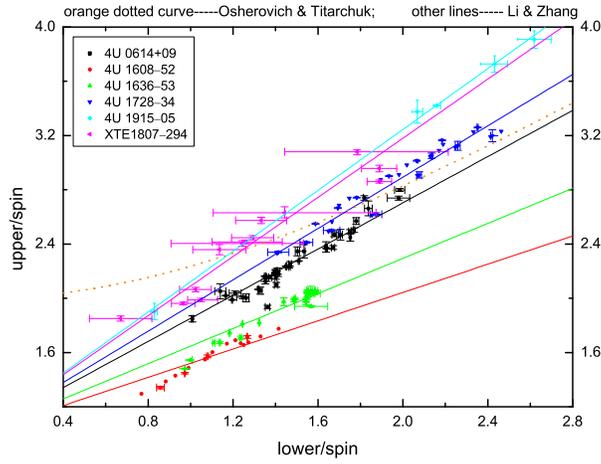}
 \end{center}
\caption{The relations between $\nu_2/\nu_s$ and $\nu_1/\nu_s$
compared with the predicted ones in Li {\&} Zhang (2005) and
Osherovich {\&} Titarchuk (1999).} \label{fig4}
\end{figure}
\end{center}

\begin{figure*}
\includegraphics[width=3.20in,height=2.35in]{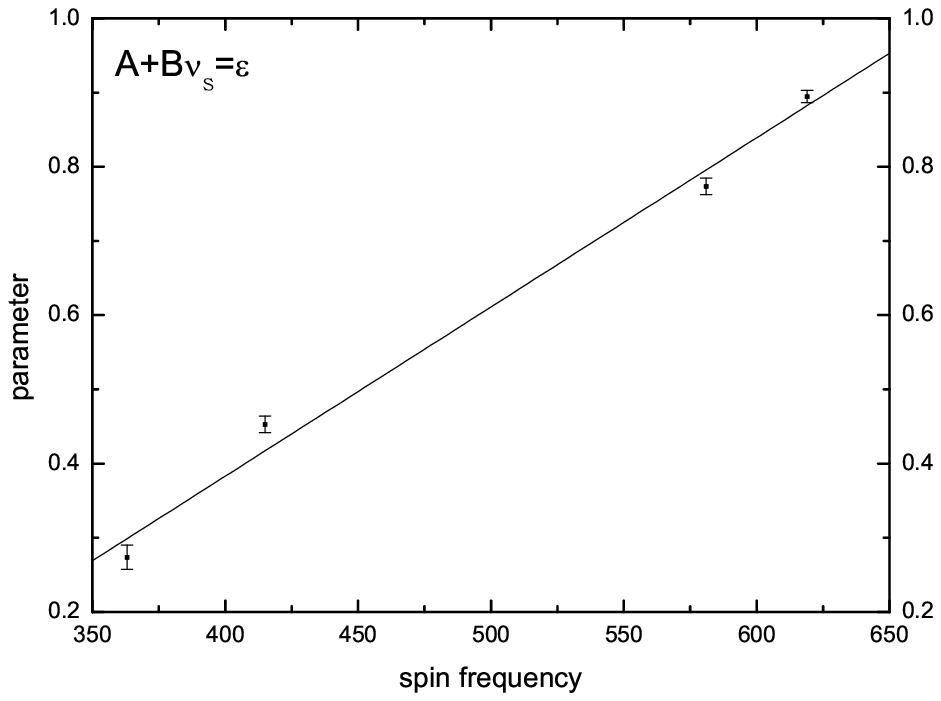}
\includegraphics[width=3.20in,height=2.35in]{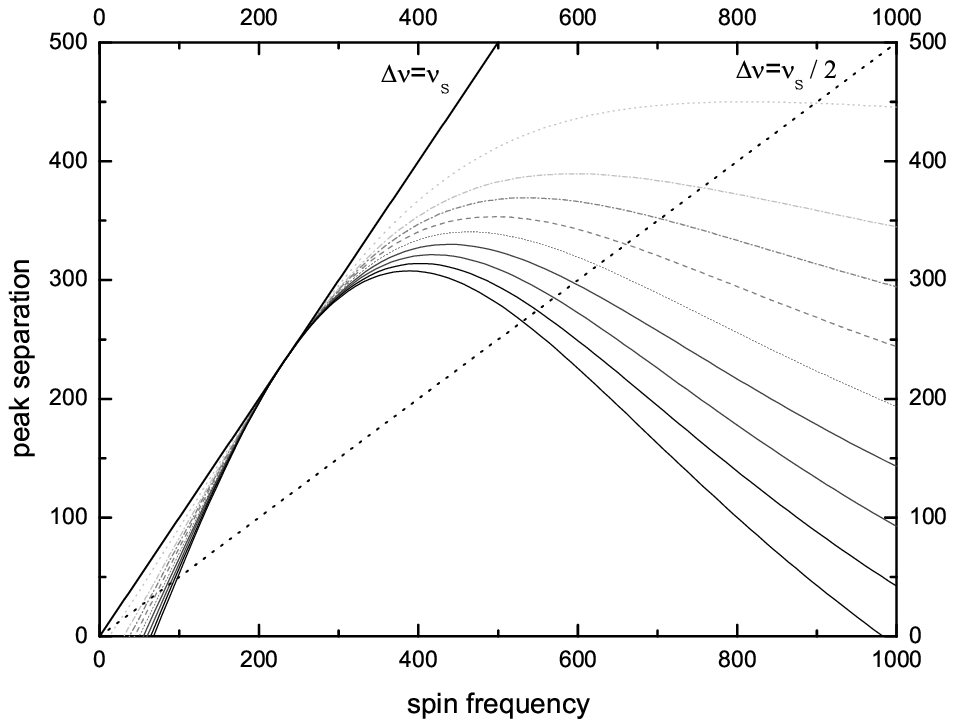}
\caption{ $Left$ The relation between the parameter $\varepsilon$
and the NS spin frequencies in the four SSCS.  $Right$ The relation
between the peak separation of the twin kHz QPOs and the spin
frequencies with different lower kHz QPO frequencies. } \label{fig5}
\end{figure*}

\clearpage
\begin{table*}
 \centering
  \caption{The measured and fitted parameters for six LMXBs.
For the first four sources we use the relation $\nu _2 = (\nu _1 +
\nu _s)/(\sqrt {1 + \varepsilon^2})$, and for the last two sources
with $\nu _2 = \sqrt {1 + \delta^2} (\nu _1 + \nu _s)$.} \label{t:3}
  \begin{tabular}{@{}l|cccc|c@{}}
   \hline
  \hline
sources  & $\nu_s$ (Hz) & $\varepsilon$ or $\delta$ & error($\pm$) & $\chi^2/DoF$ & slope\\
 \hline
4U 1728$-$34 & 363& 0.27386  & 0.01629 &\textbf{550.3/33}
&0.96449  \\
4U 1608$-$52  & 619 & 0.89454  & 0.00826  & \textbf{450.5/16}&0.74531 \\
4U 1636$-$53  & 581 & 0.77732  & 0.00768  & \textbf{551.3/24}& 0.78953 \\
4U 0614$+$09 & 415& 0.45296  & 0.01118   & \textbf{563.0/39}&0.91091  \\
  \hline
4U 1915$-$05  & 270 & 0.41657& 0.01784 & \textbf{840.3/ 4}& 1.08330  \\
XTE 1807$-$294 & 190.6 & 0.33026& 0.04213& \textbf{94.6/12} & 1.05312\\
\hline
\end{tabular}
\end{table*}
\label{lastpage}

\end{document}